\documentclass[pre,twocolumn,floatfix]{revtex4-2} 
\usepackage{float}
\usepackage{booktabs}

\usepackage{amsmath} 
\usepackage{amsfonts} 
\usepackage{graphicx} 
\usepackage{xcolor}

\usepackage{hyperref}  
\hypersetup{
    colorlinks=true,    
    citecolor=blue,     
    linkcolor=blue,    
    urlcolor=blue        
}

\usepackage{color}

\begin{document}


\title{ Extreme events in a  random set of nonlinear elastic bending waves}
\author{Murukesh Muralidhar}
\affiliation{SPEC, CEA, CNRS, Université Paris-Saclay, CEA Saclay, 91191, Gif sur Yvette, Cedex, France}

\author{Antoine Naert}
\affiliation{Laboratoire de Physique de l'\'{E}cole Normale Sup\'{e}rieure de  Lyon, CNRS,Univresit\'e Claude Bernard, 46 allée d'Italie , F-69007 Lyon, France }
\author{S\'ebastien Auma\^ itre}
\email[Corresponding author.\\ Email address: ]{sebastien.aumaitre@cea.fr}
\affiliation{SPEC, CEA, CNRS, Université Paris-Saclay, CEA Saclay, 91191, Gif sur Yvette, Cedex, France}

\begin{abstract}

\textcolor{blue}{We present an experimental setup designed to investigate the statistical properties of extreme events in random elastic bending waves induced by an electromagnetic shaker on a thin stainless steel plate. In this setup, the standard statistical criteria used to define extreme events, such as rogue waves in the sea, are not sufficiently restrictive. Therefore, we introduce a new, more restrictive criterion to quantify the occurrence of rare events, similar to those observed in wave tanks \cite{Micheletal2020}. Using this refined criterion, we explore correlations between the amplitude of extreme events and other wave characteristics, such as slopes, energy, and periods of the waves. We find that extreme events in our setup are correlated to the longest wavelength of the plate, which corresponds to the plate's mode. We also observe that the steepness and kinetic energy of these events reach their time-averaged value, as expected for these slow-varying modes of the plate. The study raises questions about the purely statistical characterization of statistically rare events and rogue waves. }
\end{abstract}

\maketitle

\section{ Introduction}
\textcolor{blue}{Beyond average values and typical fluctuations, extreme events are particularly important to study. They impose many constraints on infrastructures subject to a fluctuating environment. Among these damaging events, one can recall earthquakes, tsunamis, hurricanes, etc. All have their own statistics and definitions of extreme events, but there is a common feature: these extreme events deviate significantly from the normal distribution expectation. Nevertheless, the precise definition of extreme events is often based on the intensity and occurrence of these events and depends on their statistical properties. }\\

\textcolor{blue}{Among these extreme events, some are generated in the set of nonlinear waves. Here we will focus on extreme events in a set of elastic bending waves.  Fluid-structure interactions and nonlinear energy transfer of such mechanical waves can enhance the fundamental modes and lead to the collapse of the infrastructure  (see the famous collapses of the Tacoma Bridge \cite{Billah1991} or of the Broughton suspension bridge for examples). Since field measurements are subject to many uncontrolled external factors, it is useful to propose controlled experiments to capture the main ingredients involved in the generation of these huge events. For instance, in the case of mechanical vibration, we would like to uncouple purely linear aeroelastic effects from those involving nonlinear wave interactions in generating such dramatic events. In order to focus on the latter, we propose a very simple setup where the bending waves are generated in a thin rectangular plate forced at one end by an electromagnetic shaker. These waves, described by the F\" {o}ppl--von K\' {a}rm\'{a}n equation \cite{LandauElast}, are known to generate four-wave in interactions, implying a direct cascade of energy to small scales  \cite{During,Mordant,Arezki}. The setup becomes a canonical example of the wave turbulence study. Nevertheless, complex dissipation mechanisms invalidate the theoretical prediction \cite{During,MordantII}. To our knowledge, extreme event statistics have not been explored in such a system.}\\

\textcolor{blue}{In contrast, rogue waves at the sea surface are an example where extreme events in nonlinear waves have been abundantly studied either from field measurements \cite{FreakWavesBook, AnnRevRogWave,Toffoli_Onorato} or in well-controlled experiments in wave tanks \cite{ChabchoubExp,OnoratoExp,Micheletal2020}. In the following, we will use these studies as guidelines to define extreme events in our setup. Since bending waves and surface waves share some common features (both are two-dimensional nonlinear waves with four-wave interaction) but also many differences (e.g., the number of interacting waves is conserved for surface waves but not for bending waves;  hence, the former exhibits an inverse cascade of wave action, whereas the latter does not). A comparative analysis of both systems may provide valuable insights into the underlying mechanisms involved in the formation of extreme events.}\\

We first briefly present the experimental setup for bending waves. Then, we characterize the statistical properties of the local displacement, especially the fraction of extreme events, for various forcing amplitudes and frequencies. To capture the underlying mechanism, we measure the correlation between the amplitudes of these rare events and different characteristics of the waves (periods, steepness, energy). The concluding part proposes a scenario for the generation of rare events in a set of bending waves and compares it  with that in gravity surface waves \cite{Micheletal2020}.

\section{Experimental Device and data processing}

\begin{figure}
\resizebox{.35\textwidth}{!}{%
\includegraphics{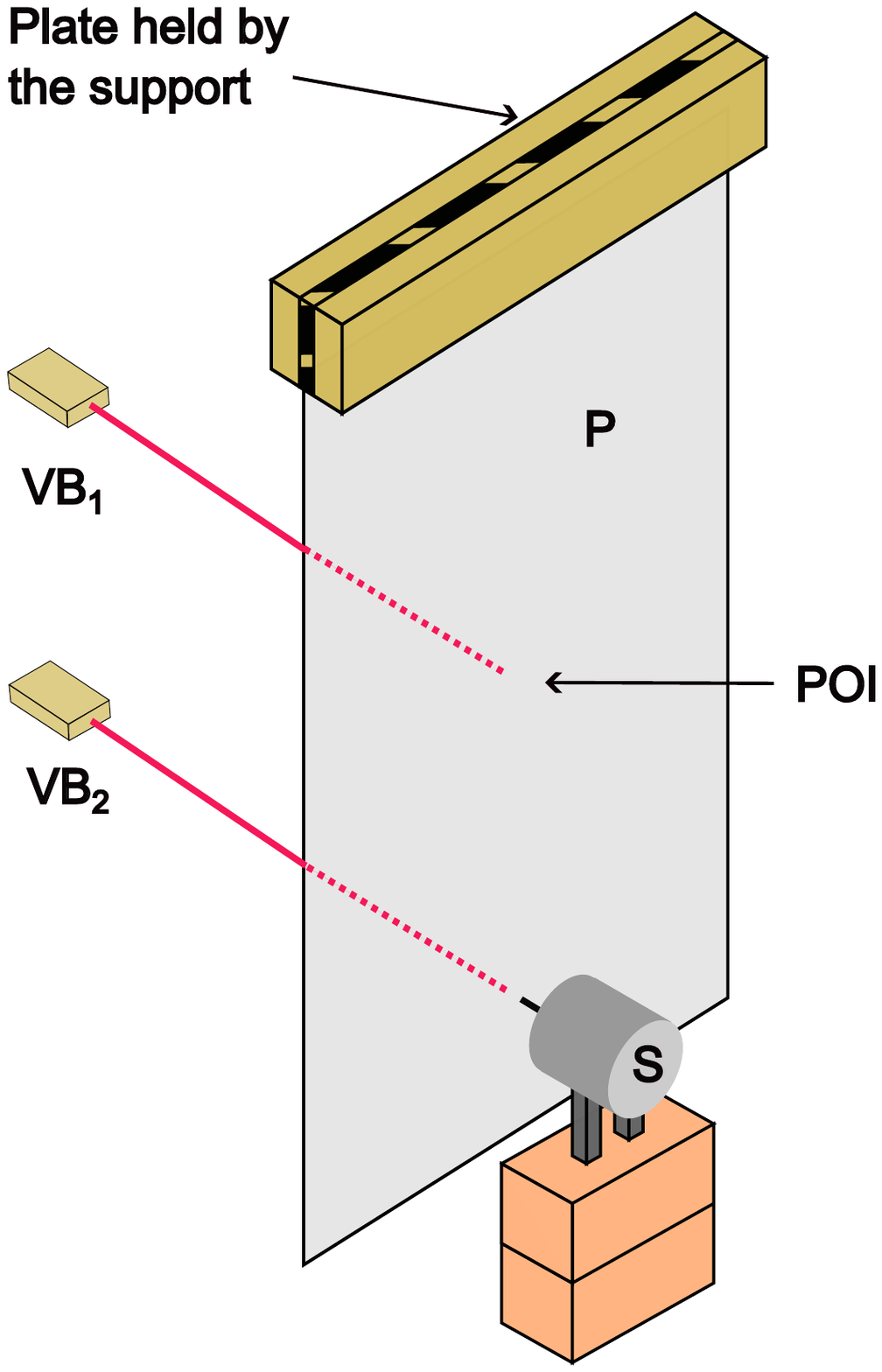}
}
\caption{Experimental setup: a thin stainless steel plate, P (2000 $\times$ 1000 $\times$ 0.5 mm$^3$ is forced by an electromagnetic shaker, S, at frequency $f_d$. A vibrometer, VB$_1$ (placed behind the plate), measures the transverse displacement of the plate at the point of interest, POI, in the middle of the plate. A second vibrometer, VB$_2$ (also placed behind the plate), is used to measure the velocity at the injection point. }
\label{ExpSetup}
\end{figure}

The setup, shown in Figure \ref{ExpSetup}, is similar to the one presented in \cite{AumaitreNaertMiquel}. A thin elastic plate ($2L\times L\times l=2000\times 1000\times 0.5$ mm$^3$) is fastened to its support at the top and to an electromagnetic (EM) shaker attached at 10 cm from the bottom. Other boundaries are free. In our device, the bending waves follow a quadratic dispersion relation given by $f=c.lk^2$ with $k=2\pi/\lambda$ the wavenumber, $\lambda$ as the wavelength and where $c=\sqrt{\frac{E}{12(1-\sigma^2)\rho}}$ has the dimension of a velocity ($E$ being the Young's modulus,  $\sigma$ the Poisson ratio, and $\rho$ the specific mass of the stainless steel). \textcolor{red}{We are going to assume in the following that the dispersion relations hold regardless of the amplitude of the vibration. We base this hypothesis on \cite{MordantII}  where only a slight shift of the dispersion relation is observed under a strong forcing of the same device. The relevance of this assumption for extreme events is discussed in the last section}. 

The EM shaker is used to drive the plate into an out-of-equilibrium steady state by imposing a sinusoidal motion of varying frequency and amplitude. \textcolor{red}{We impose the voltage applied to the coil of the shaker, which drives the motion of a permanent magnet (similar to a loudspeaker). This is equivalent to fixing the applied force on the moving magnet. A force sensor placed on the rigid link between the shaker and the plate shows almost sinusoidal fluctuations. The velocity (and hence the current in the coil) fluctuates, reflecting the feedback of the plate on the shaker. The resulting fluctuations of injected power have been studied in \cite{Cadot_2008,Apffel_2019}.} The driving frequency $f_d$ is such that the associated forcing wavelength, $\lambda_o$, is smaller than the plate size, $\lambda_o < L$, in order to capture the role of the waves at scales larger than those $\lambda_o$, which are at equipartition of energy \cite{AumaitreNaertMiquel}. Additionally, forcing bending waves at higher frequencies, such as 150 Hz and 250 Hz, as in our study, offers a distinct advantage by providing significantly more statistical data in a shorter time compared to surface wave experiments, which are typically limited to 2 Hz. A vibrometer is used to measure the plate displacement. Measurements are taken around the plate's center. \textcolor{blue}{ One must mention that the statistical  properties are equivalent  once normalized by the standard deviation as long as we remain away from the injection point or the boundaries. All measurements are taken 20 cm from the boundaries}. In addition, we measure the velocity and the force at the energy injection point in order to quantify the level of turbulence based on the energy input. For each forcing parameter, we performed a 10-hour measurement at a sampling rate of 4096 samples per second. The convergence of the extreme events statistics was tested with a 62-hour run. 

\begin{figure}
\resizebox{.39\textwidth}{!}{%
\includegraphics{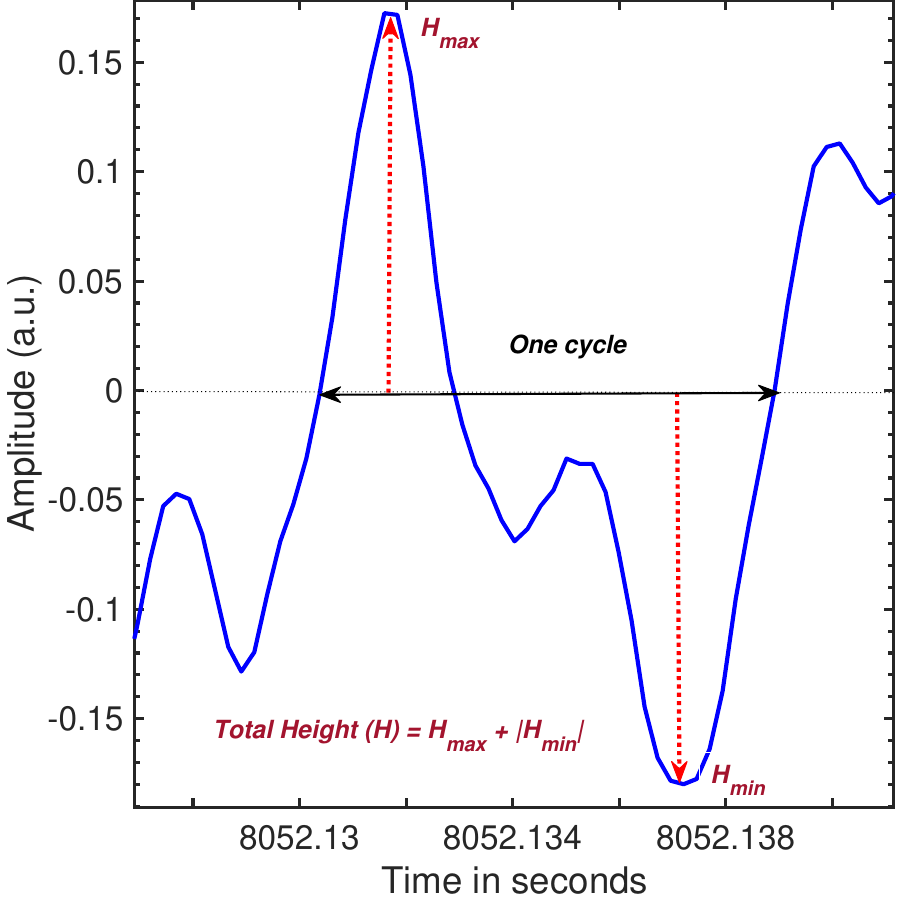}
}
\caption{The temporal trace of a typical transversal displacement signal zoomed around two zero crossings with the same slope sign. It allows us to define the local wave period $T_i$, the local maximum, $H_{max}$, and minimum, $H_{min}$, over such period, and the total height $H=H_{max}-H_{min}$.}
\label{ZoomAmp}
\end{figure}

To mimic the rogue wave analysis, as done for surface waves, we define a {\it local wave period} $T_i$ as the $i$th interval between two zero crossings with the same slope sign of the transverse displacement $\eta$. The wave crests ($H_{\text{max}}$) and wave troughs  ($H_{\text{min}}$) are defined as the maximum and minimum values, respectively, during the local period (see the temporal trace in Figure \ref{ZoomAmp}). Then, the wave height, $H$, is defined as the sum of $H_{max}$ and $|H_{min}|$ . 

\section{experimental results}
\subsection{Statistics of the the bending waves amplitudes}
\subsubsection{Statistical properties of the displacement} 

\begin{figure}
\resizebox{.4\textwidth}{!}{%
\includegraphics{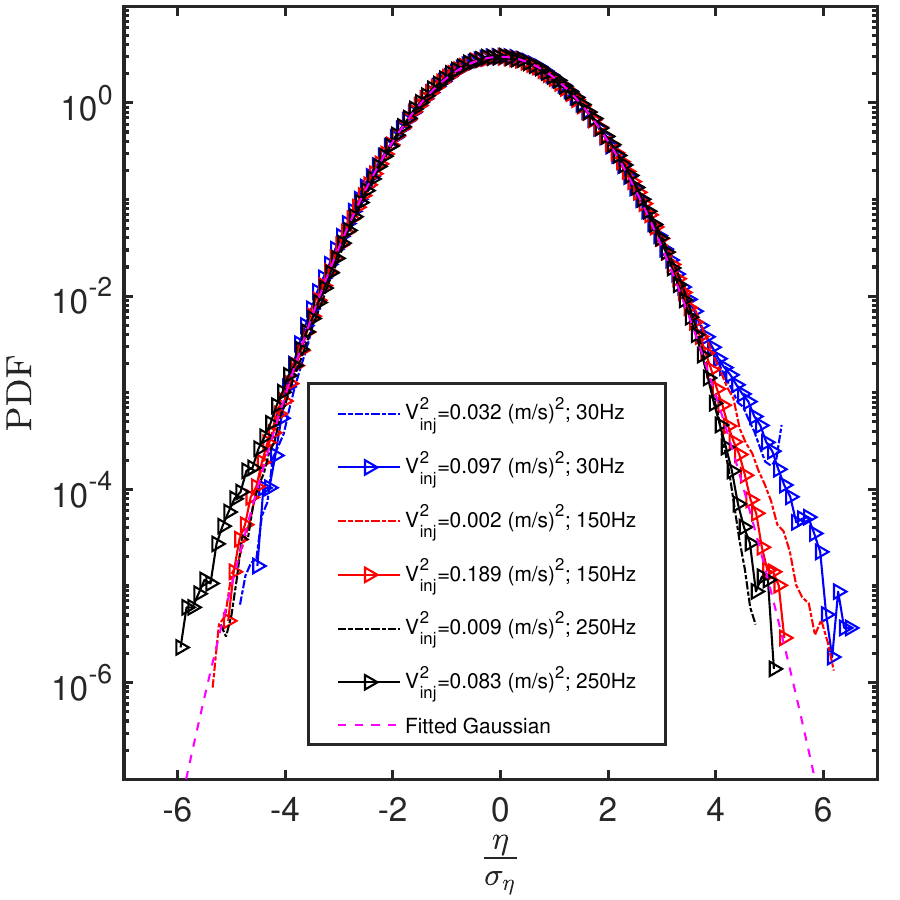}
}
\caption{Probability Density Functions of transversal displacement normalized to its standard deviation for different forcing amplitudes and frequencies. A Gaussian fit (in dotted magenta) is plotted for comparison. }
\label{norm_pdf_disp}
\end{figure}

\begin{figure}
\resizebox{.4\textwidth}{!}{%
\includegraphics{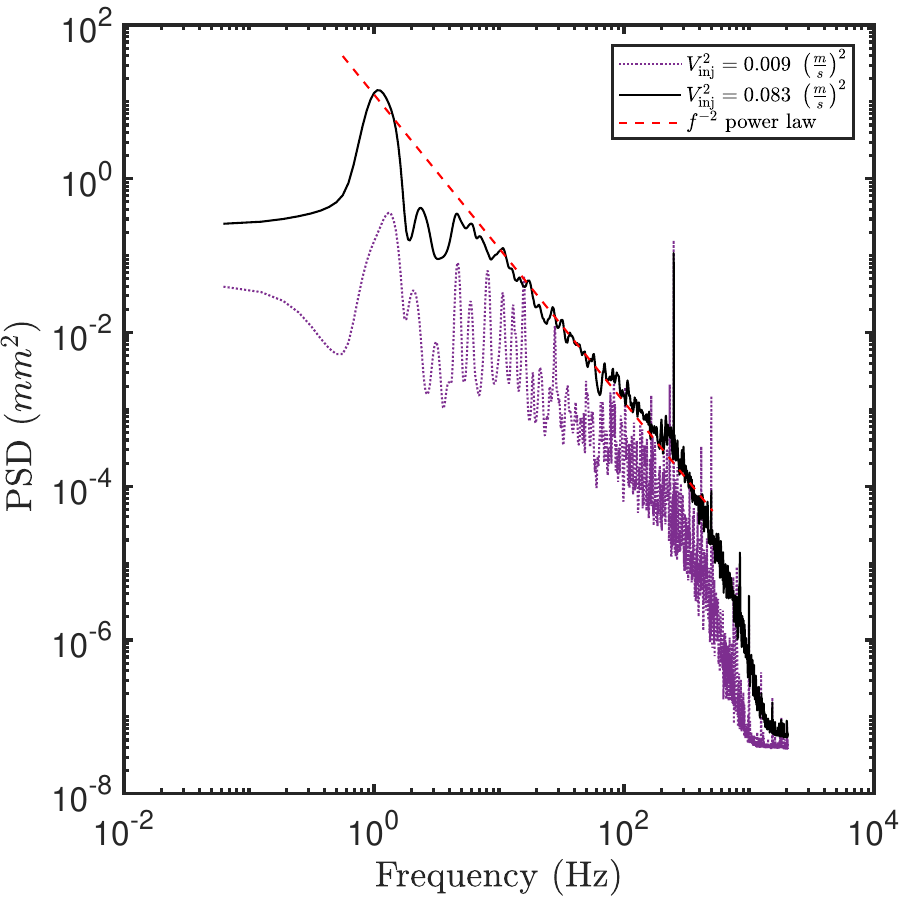}
}
\caption{Power Density Spectra of the transversal displacement at forcing frequency $f_d=$250 Hz and at small forcing amplitude  $V_{inj}^2=0.009 (m/s)^{2}$  (in dotted purple) and large forcing amplitudes $V_{inj}^2=0.083 (m/s)^{2}$  (in black). The dot-dashed red line represents the power law $f^{-2}$ expected for the equipartition regime.}
\label{psdDisp}
\end{figure}

We first briefly consider the statistical properties of the transverse displacement $\eta$, which is directly measured by the vibrometer at a point near the plate's center. In the nonlinear regime, the typical displacement fluctuations are represented by their Probability Density Functions (PDFs) in Figure \ref{norm_pdf_disp} and their Power Density Spectrum (PDS) in Figure \ref{psdDisp}. In Figure \ref{norm_pdf_disp}, fluctuations are nearly Gaussian in the nonlinear regime, sometimes exhibiting a small skewness. This skewness is very sensitive to even tiny misalignment of the plate with the vertical axis. The PDS are shown in Figure \ref{psdDisp} at low forcing and in the nonlinear regime with a forcing frequency $f_d=250$ Hz. The low-forcing spectrum exhibits the forcing frequency and the modes of the plates, whereas the spectrum in the nonlinear regime is continuous above and below $f_d$ (extending up to a peak near 1 Hz). The power law $f^{-2}$ observed for $f \leq f_d$, is consistent with an equipartition of the energy, with no energy flux on average in this frequency range \cite{AumaitreNaertMiquel}. Note that this range becomes significantly narrower when the forcing frequency is set to  $f_d=30$ Hz.
\subsubsection{Statistical properties of the wave height}

\begin{figure}
\resizebox{.4\textwidth}{!}{%
\includegraphics{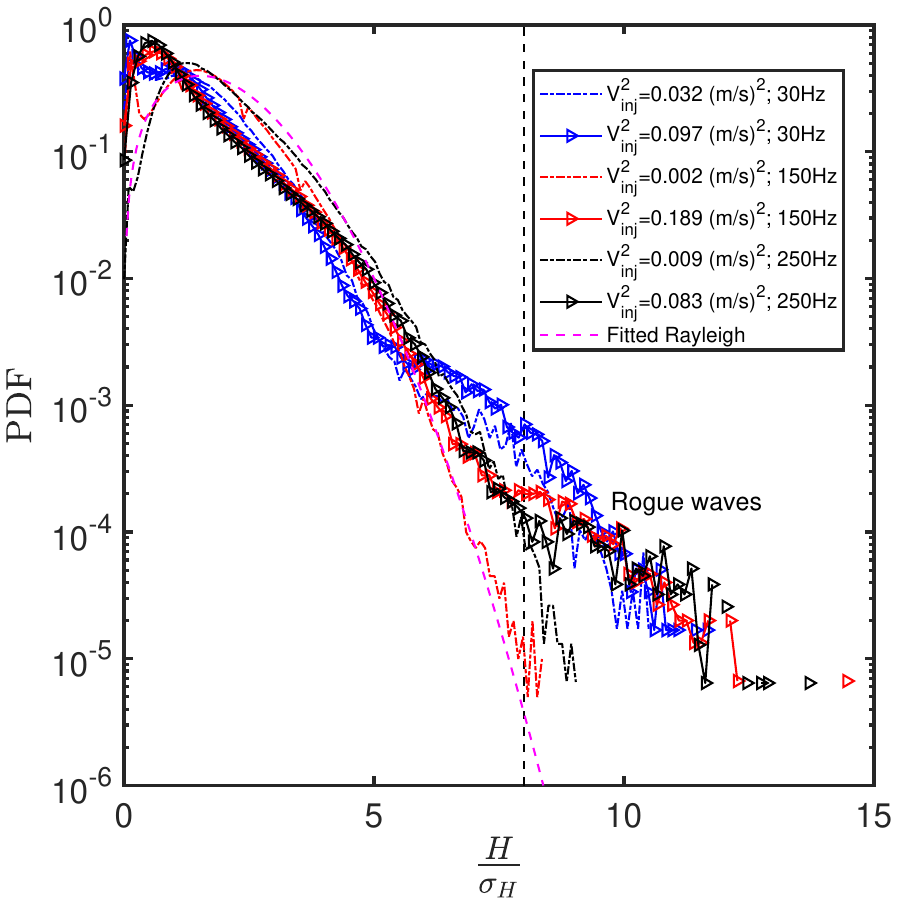}
}
\caption{PDF of normalized total height distribution $H$ for different forcing amplitudes and frequencies with the dashed black line representing the region where criteria for rogue events are met.}
\label{norm_pdf_total_height}
\end{figure}

\begin{figure}
\resizebox{.4\textwidth}{!}{%
\includegraphics{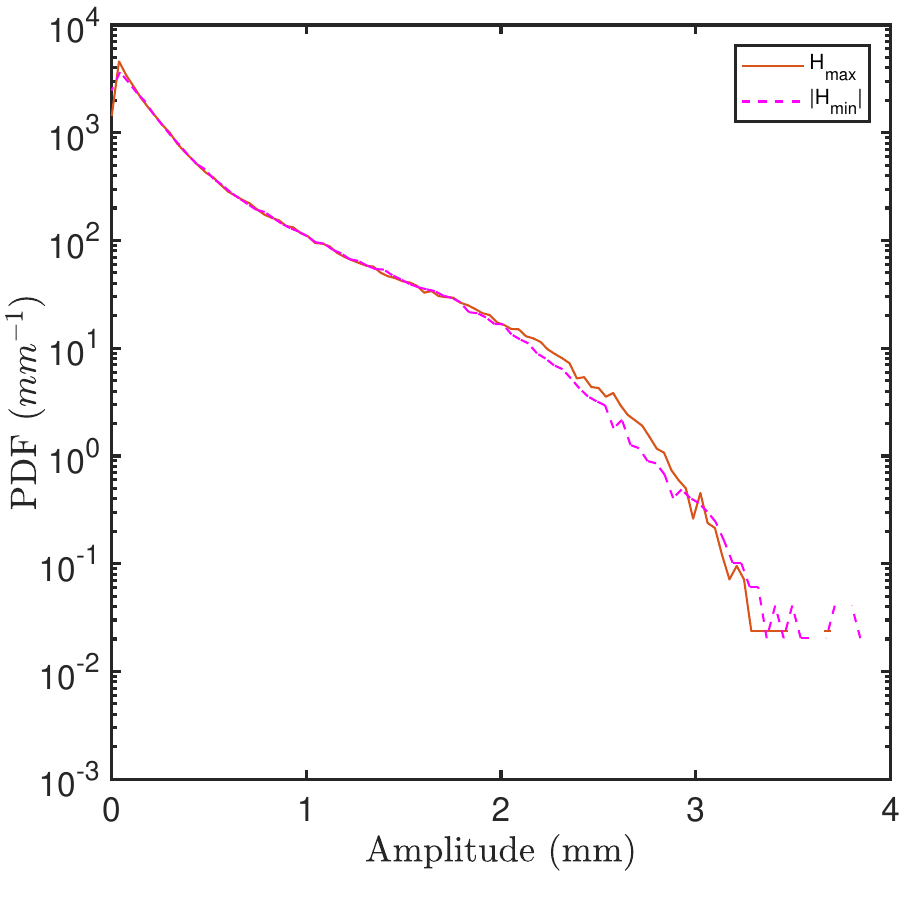}
}
\caption{PDF of the extrema of the transversal displacement for one typical measurement ($V^2_{rms,inj}=0.083 (m/s)^{2}$, $f_d=250$ Hz). }
\label{HminHmaxPDF}
\end{figure}

We now focus on the height $H$, as previously defined. The PDF of the height is shown in Figure \ref{norm_pdf_total_height}. Although the PDF of the transversal displacement is quite Gaussian, the shape of the PDF of the height is far from the Rayleigh distribution. Such a Rayleigh distribution would describe the height of a random set of Gaussian and uncorrelated transversal displacements. It also accurately describes the height of gravity waves on water surfaces. The PDFs of the bending wave height have fatter tails, amplifying the occurrence of rare events in the nonlinear regime. In contrast to surface waves, these peculiar shapes are also found in the PDF of $H_{max}$ and $H_{min}$ which are symmetrical in Figure \ref{HminHmaxPDF}, as expected by the symmetries of the setup (i.e., the fluctuations of $H_{max}$ and $H_{min}$  seem poorly affected by the small skewness observed for the displacement). 

The traditional definition of rogue waves involves the significant wave height, \( H_s \), which is the mean wave height of the highest third of the wave field. The rogue waves have a height such that \( H \geq 2H_s \). For surface sea waves, the height distribution follows a Rayleigh distribution and \( H_s \) corresponds to four times the standard deviation of the surface elevation, \( \sigma_e \) \cite{Longuet_Higgins,ChapBook}. Therefore, the most widely accepted definition of rogue waves is waves with a height \( H \) such that \( H \geq 8\sigma_e \). 

In contrast, bending elastic waves exhibit a distribution with much fatter tails (see Figure \ref{norm_pdf_total_height}). Consequently, using the criterion \( H \geq 2H_s \) results in an event rate (about 1\%) that is too high to classify such waves as rare. On the other hand, the stricter criterion \( H \geq 8\sigma_e \) is overly restrictive, as it corresponds to only one or two events occurring during 10 hours of experiments where such events are observed. This equates to a fraction of rare events less than \( 5 \times 10^{-4}\,\% \) in strongly forced cases, which is significantly lower than the rogue wave occurrence rates measured in wave tanks.

\begin{figure}
\resizebox{.4\textwidth}{!}{%
\includegraphics{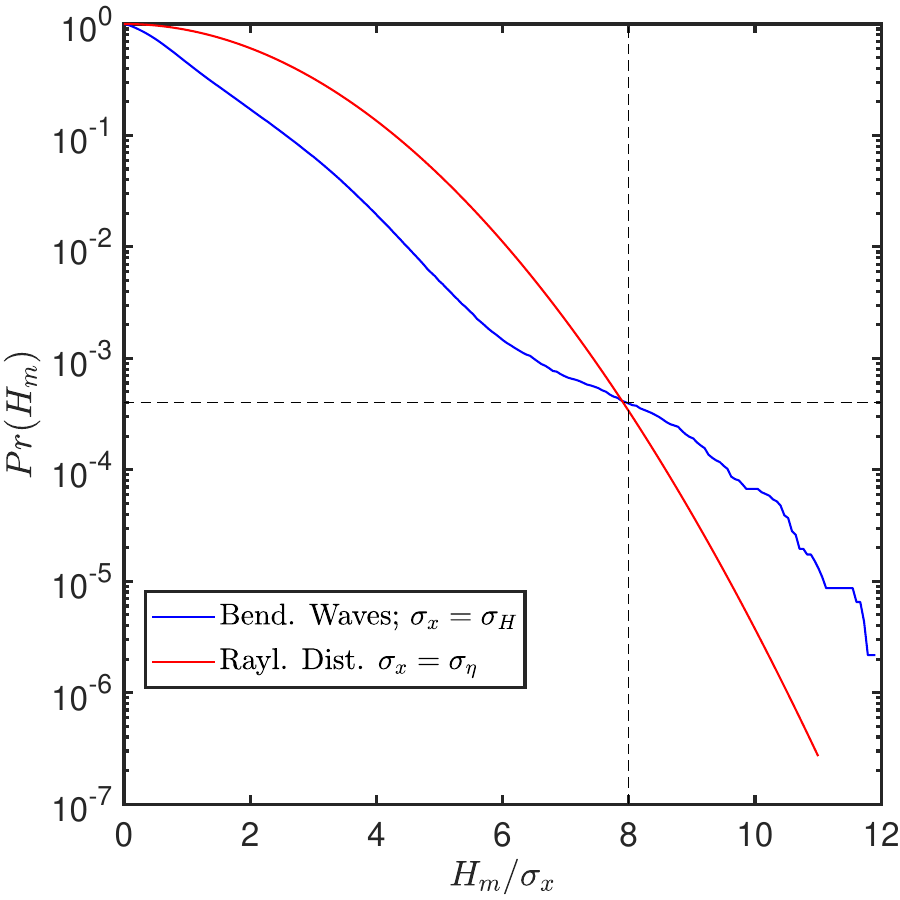}
}
\caption{\textcolor{blue}{Probability of exceedance, $Pr(H_m)$ of the bending waves height normalized by the standard deviation of the waves height, $\sigma_\eta$, (blue) and of waves following the Rayleigh distribution normalized by the waves elevation $\sigma_\eta$. The forcing frequency of the bending waves is 150 Hz and and a standard deviation of the velocity at the injection point of $V_{rms,inj}0.302$ m/s}}
\label{ProbOcc}
\end{figure}

\textcolor{blue}{This discrepancy is well illustrated in Figure \ref{ProbOcc} where we compare the probability of exceedance for a wave height following a Rayleigh distribution and for bending waves in the non-linear regime. The probability of exceedance $Pr(H_m)$ is given by the probability that the height exceeds a given value $H_m$. Bending waves differ strongly from the probability of exceedance expected for a Rayleigh distribution. In figure \ref{ProbOcc}, we normalize $H_m$ by $\sigma_\eta$ the standard deviation of the elevation for the probability of exceedance of the Rayleigh distribution. $H_m$ is normalized by $\sigma_H$ the standard deviation of the height in the case of bending waves. Figure \ref{ProbOcc} clearly shows that the probability to exceed $H\geq 8\sigma_\eta$ for rogue waves following the Rayleigh distribution, almost corresponds to probability to exceed $H\geq 8\sigma_H$ for the bending waves. Therefore, we choose the latter criteria to define the extreme events in our set of bending waves. It allows us to recover the same fraction of rare events as in \cite{Micheletal2020}}

\begin{figure}
\resizebox{.4\textwidth}{!}{%
\includegraphics{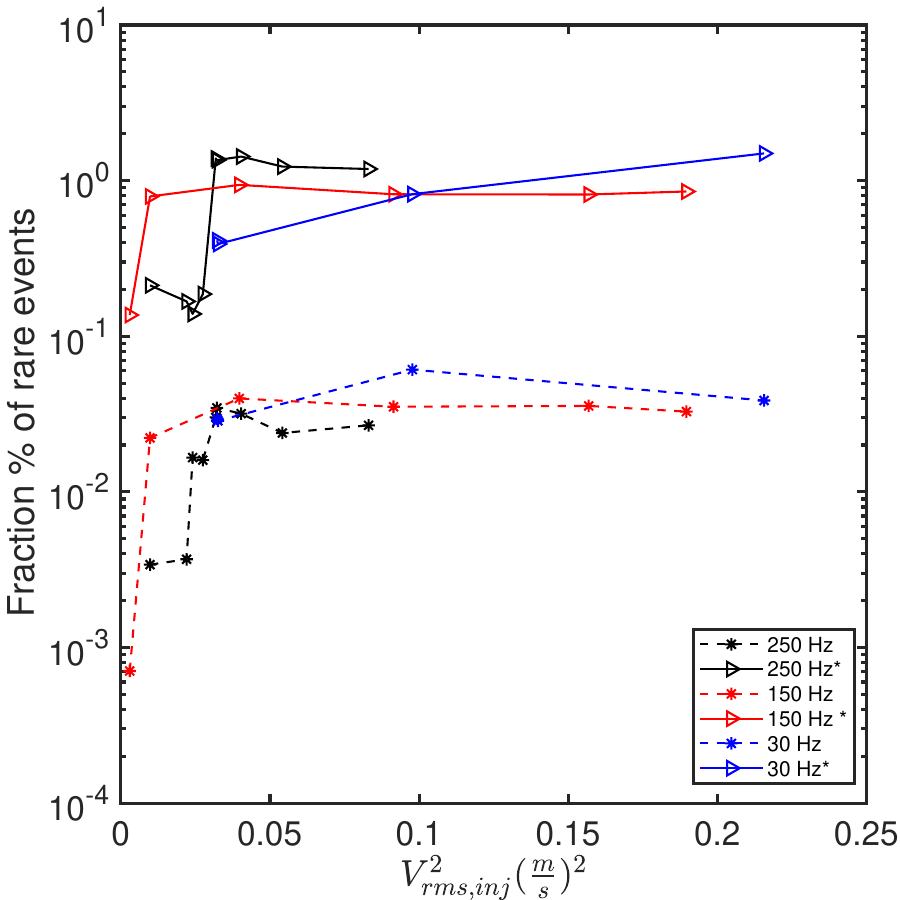}
}
\caption{Fraction \% of extreme events vs the squared injection velocity for different forcing frequencies.$"*"$ corresponds to the fraction percentage of extreme events observed if using a significant height based definition, which corresponds to approximately $1\%$ of the total recorded events.}
\label{ExtremEvents}
\end{figure}

 These fractions of rare events are shown in Figure \ref{ExtremEvents} for various forcings. In the strongly forced fully nonlinear regime, it converges around a fraction of 0.02 to 0.06 \% with no clear dependencies on the forcing frequency. This corresponds to a few hundred events during the 10 hours of measurements. In the next section, we look for the correlation of these rare events with other characteristics of the waves. \textcolor{blue}{One must insist that statistical properties shown in figures \ref{norm_pdf_total_height}--\ref{ExtremEvents} are insensitive to the precise position of the measurement point. Only the onset of the asymptotic non-linear regime could depend on it.}

\subsection{Correlation}
\begin{figure}
\resizebox{.4\textwidth}{!}{%
\includegraphics{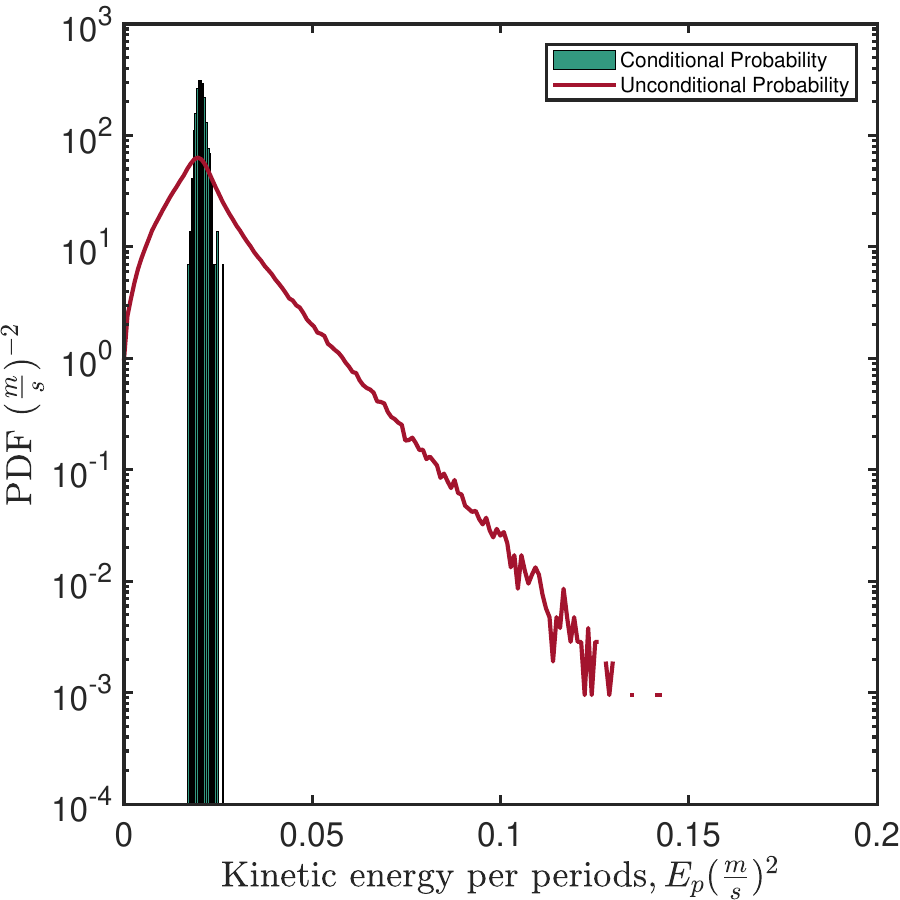}
}
\caption{Probability Density Function of the kinetic energy per unit mass during a local period without any condition (in dark red) and conditioned to the presence of a extreme event during the local period (teal bar histogram)}
\label{PDF_energy}
\end{figure}

To capture the correlations between extreme events and other quantities characterizing the wave, like its kinetic energy per unit mass, its slope, or its local periods, we compare the PDF conditioned to the presence of an extreme event within the local period with the PDF without any condition. If the extreme events were uncorrelated to the considered quantities, the PDF would remain unchanged. We focus on the driving in the strong nonlinear regime, where the number of extreme events reaches its asymptotic values (i.e., for a forcing energy $V^2_{rms,inj}$ around 0.05 $(m/s)^{2}$ in Figure \ref{ExtremEvents}). Due to the destructive potential of rogue waves at the sea surface, we might assume that they contain a lot of energy. In the case of our bending waves, we compare in Figure \ref{PDF_energy} the PDF of the kinetic energy per unit mass during a local period conditioned to the presence of extreme events during this local period and the unconditioned PDF. The width of the conditional probability is shrunk by about a factor of 7, but the average is about the same as the unconditional average (the ratio of the conditioned to the unconditioned average kinetic energy per unit mass is about 0.95 in the case of Figure \ref{PDF_energy}) i.e all periods containing rare events have almost the averaged kinetic energy per unit mass.
 
\begin{figure}
\resizebox{.4\textwidth}{!}{%
\includegraphics{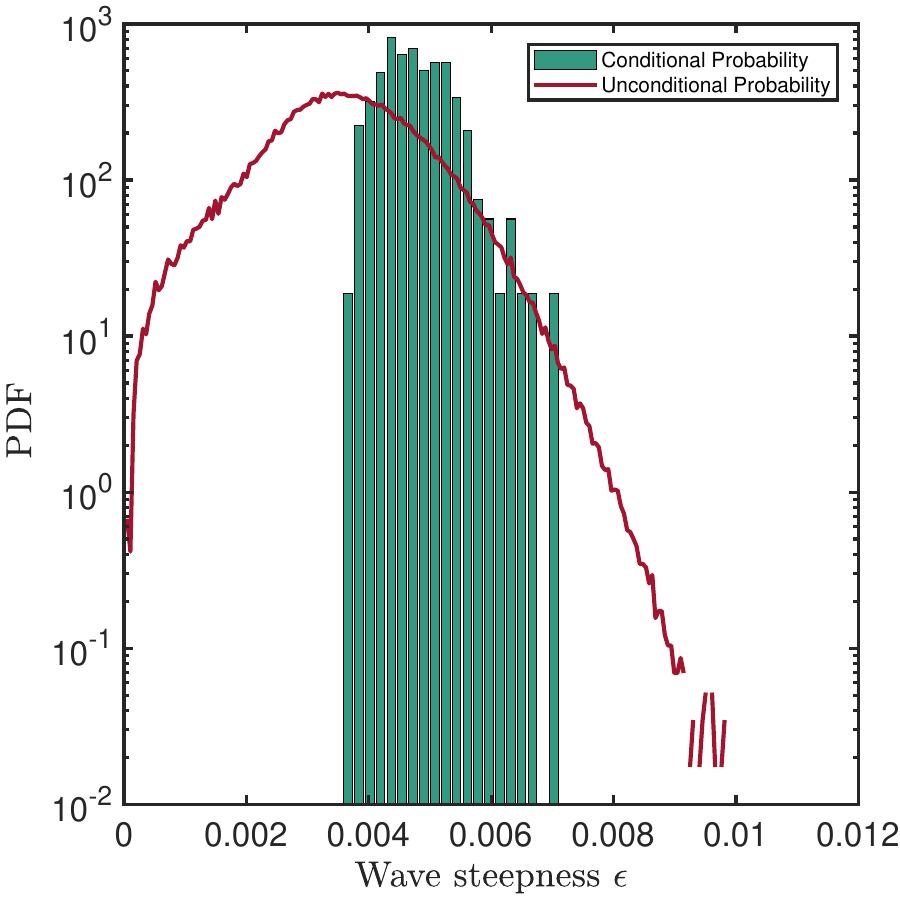}
}
\caption{Probability Density Function of the wave steepness without any condition (in dark red) and conditioned to the presence of a extreme event during the local period (teal bar histogram)}
\label{PDF_steepness}
\end{figure}

Rogue waves in the ocean are sometimes described as walls of water. This suggests that rogue waves steepness is large. We probe the correlation between extreme event amplitudes and the wave steepness of the bending waves. The steepness $\epsilon$ is the ratio of the wave amplitude $H_i$ over half the wavelength, $\lambda/2$. In terms of the local wave period and height, it can be recast as: $\epsilon=\frac{2H_i}{\sqrt{2 \pi c lT_i}}$, where $H_i$ is the height corresponding to the local period $T_i$ as defined previously. Here, we used the dispersion relation of the bending waves, $\frac{2\pi}{T}=c \cdot l\left(\frac{2\pi}{\lambda}\right)^2$ which is assumed to remain valid in the nonlinear regime of WT \cite{MordantII}. Figure \ref{PDF_steepness} shows the PDF of the steepness $\epsilon$ when rare events occur and compares it to the unconditional steepness PDF. Here again the fluctuations of the rare events steepness are shrunk (about a factor of 2), but their average is higher, about 30 \%. The fact that the steepness of rare events remains of the same order as that of other waves is inconsistent with the "wall of water" image often derived from sailors' testimonies, which implies extreme steepness for rogue waves. However, it is important to note that this intuitive picture is not always accurate. Steepness alone may not reliably indicate rogue events, as demonstrated by studies of field measurements from real sea states, which show that wave fields without rogue waves can still exhibit high wave steepness. \cite{FieldMeasurementsofRogueWaterWaves}

\begin{figure}
\resizebox{.4\textwidth}{!}{%
\includegraphics{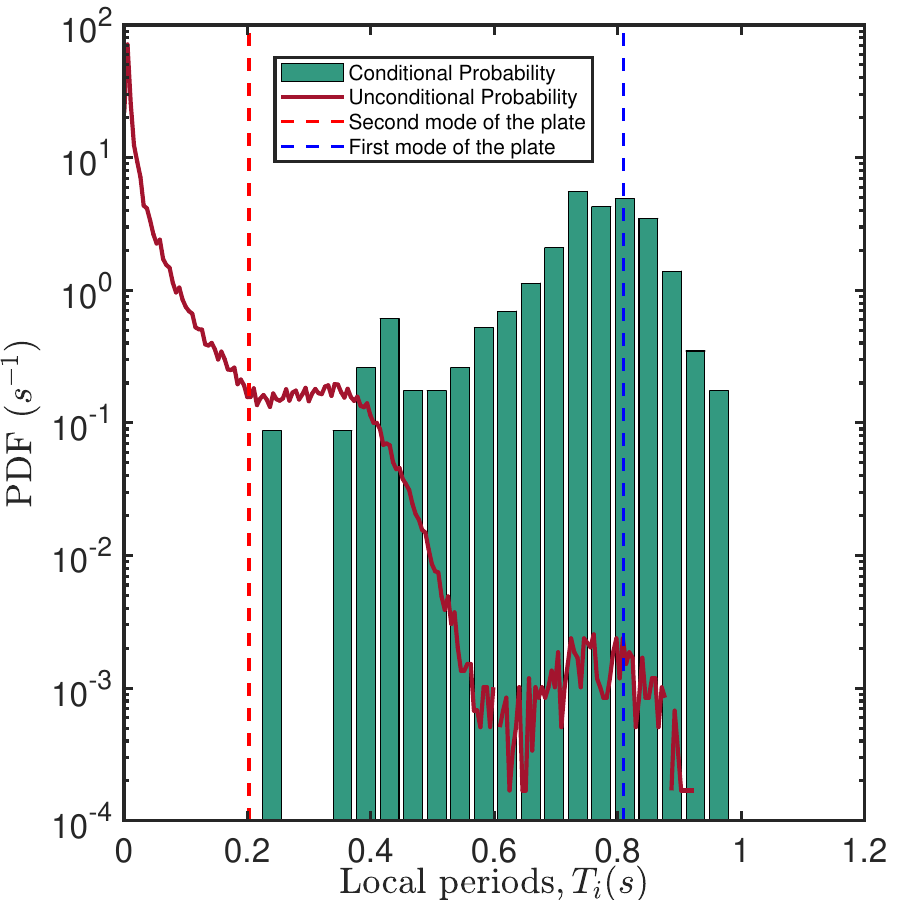}
}
\caption{Probability Density Function of the  local period without any condition (in dark red) and conditioned to the presence of a extreme event during the local period (teal bar histogram)}
\label{PDF_period}
\end{figure}

Finally, we compare in Figure \ref{PDF_period} the statistics of the local period of rare events to the unconditional local period statistics. Here the local periods of extreme events are clearly concentrated around the highest periods that coincide with a bump in the unconditional PDF of the local period. These local periods around 0.8 s correspond to nearly a 2 m wavelength. It is one wavelength along the length and half a wavelength along the width of the plate. Actually, this fundamental mode is clearly visible in the spectra in Figure \ref{psdDisp}. 

All these observations made in the case with $V^2_{rms,inj}=0.083 (m/s)^{2}$ and $f_d=250$ Hz in Figures \ref{PDF_energy} -- \ref{PDF_period}, remains true for all forcing at 150 and 250 Hz,  as long as we are in the nonlinear regime.
At 30 Hz, where the equipartition stage is reduced, the steepness of rare events is slightly higher. We report in the table in appendix A: the ratio of the unconditioned statistics over the conditioned one for the kinetic energy per unit mass during a period, for the steepness and for the local period for the average and the variance of all our experimental runs.

\section{Discussion $\&$ Conclusions}
Based on these findings, we can elaborate the following scenario for the generation of rare events in a set of bending waves. Although an inverse cascade is not expected in bending wave turbulence, there is at least an equipartition of the energy per mode at wavelengths larger than the forcing scales in the stationary state. This is enough to bring energy to the fundamental modes of the plate. This large wavelength does not require a high steepness to reach a very large height. The steepness must be slightly higher when the equipartition range is reduced.  \textcolor{red}{The fact that the steepness remains moderate for extreme events supports the assumption on the validity of the dispersion relation even for these extreme events. Indeed, the steepness is the relevant factor to probe the strength of nonlinear effects. We expect a strong departure from the dispersion relation only if nonlinearities are strong}. The extreme events develop over the fundamental modes of the plate. Since these rare events occur over the longest wavelength, their kinetic energy per unit mass during these modes reaches its time average value.

\textcolor{blue}{Our experiment shows that the mechanisms generating extreme events in a set of bending waves differ from those observed on the sea surface. Indeed, the latter can be reduced to the  Nonlinear Schr\"odinger equation when interactions are weak and the spectrum is narrowband.  In one dimension, a modulational (or Benjamin--Feir) instability has been demonstrated theoretically and observed experimentally in a long flume \cite{ONORATO2006586,Bonnefoy_2020,Redor_2021}. This mechanism is a strong candidate to explain rogue waves in a sea state. It is quantified by the Benjamin-Feir index $BFI=2\epsilon /(\Delta k/k_o)$ , defined as twice the wave slope over the relative bandwidth of the spectrum. The bandwidth is given by $\Delta k/k_o=\Delta f/(2f_o)$ where $k_o$ (respectively, $f_o=m_1/m_o$) is the main peak of the spectrum and $\Delta k$ (respectively, $\Delta f= f_o\sqrt{m_om_2/m_1^2-1}$) is its corresponding bandwidth. Here, $m_i=\int f^iP(f) df$  denotes the ith moment of the power spectral density $P$.  For the bending waves, we find that BFI is smaller than 0.1, allowing us to discard modulational instability. Furthermore, in two dimensions, there is no clear evidence of the  modulational instability. Moreover, field measurements show that rogue waves can develop even when $BFI < 1$ \cite{Toffoli_Onorato} and the observation of some rogue waves cannot be explained by NLS \cite{Gemmrich_Cicon}. It implies that a high BFI may not be a necessary condition for the occurrence of rogue waves. } 

\textcolor{blue}{A fundamental difference comes from the spectra of both wave fields. Usually, the sea spectrum is much narrower than the spectra plotted in Figure \ref{psdDisp}. The fundamental modes of the elastic plate play a major role, and the rare events correspond to the longest wavelength according to our statistical definition of the extreme events. This correlation has not been reported for surface waves. In contrast, the rogue waves occur near the Tayfun period \(1/f_T\), where the Tayfun frequency is \(f_T = \tilde{f_o} / \left[ 1 + (\Delta f/f_o)^2 \left(1 + (\Delta f/f_o)^2\right)^{-3/2} \right]\)\cite{Tayfun2,FreakWavesBook,Micheletal2020}. This raises a question: in our setup, there is no inverse cascade, yet energy equipartition between modes at low wavelength is enough to populate the fundamental mode of the plate. In surface waves, where an inverse cascade should enhance the lowest mode, the inverse cascade shrinks to a small range. What are the dissipative phenomena that shrink the inverse cascade and limit the growth of the first mode in wave basins?}

\textcolor{blue}{This study leverages the advantages of bending elastic waves, where we observe extreme states with an occurrence similar to those in surface gravity waves but driven by different mechanisms. By using forcing frequencies nearly 100 times higher than those in surface gravity wave experiments, we can collect much more statistical data in a shorter time. This allows us to get statistical properties of the wave conditioned by the presence of extreme events. This provides valuable input for understanding the underlying mechanisms driving these extreme events.\\}

\textcolor{red}{
One main drawback is that these measurements are limited to single-point data. Although our study provides valuable insight, wave phenomena are inherently spatio-temporal. Point measurements lack the spatial resolution needed to understand many features like energy redistribution or wavefront evolution. Hence, our interpretations of the correlations might fail to capture the complete picture. There are many spatio-temporal measurement techniques that could be applied on the plate to capture full wave dynamics, such as the FTP method \cite{MordantII}. However, simply recording indefinitely would require an immense amount of data storage and processing power, let alone successfully capturing the moment when rare events occur. Using event-based cameras or scanning laser vibrometers could offer possible solutions. }


\begin{acknowledgments}
We wish to acknowledge the ANR for financial support. We would also like to thank G. Michel, B. Gallet, C. Touzé, and S. Varghese for fruitful discussions. 
\end{acknowledgments}

\appendix
\section{TABLES}
\begin{table*}[htbp]
\caption{\label{tab:table1}Forcing frequency 30 Hz}
\begin{ruledtabular}
\begin{tabular}{ccccc}
 &\multicolumn{2}{c}{}&\multicolumn{2}{c}{}\\
   &0.5Vpp&0.5Vpp(LONG)\footnote{Long measurement of 62 hours to check convergence\\}&1.0Vpp
&1.5Vpp\\ \hline 
 $\eta_{\text{rms} (mm)}$\footnotetext[0]{  $\eta_{\text{rms}}$ is the r.m.s transversal displacement; $V_{\text{rms,inj}}$ is the r.m.s injected velocity;  K is the kurtosis; S is the skewness;}&$1.159$&$1.158$ &$2.230$& $3.596$\\
 $V_{\text{rms,inj} (\frac{m}{s})}$ 
 &$0.179$&$0.180$&$0.312$&$0.464$\\
 K\footnotetext[0]{ $N_{tot}$ is the total number of all recorded events;  $N_{re}$ is the number of detected rare events; frac is the ratio of  $\frac{N_{re}}{N_{tot}}$} &$3.057$&$3.010$&$3.105$&$3.057$\\
 S&$0.049$&$0.016$&$0.001$&$-0.111$\\
 $N_{tot}$
 & $5.484\times10^5$&$3.454\times10^6$&$5.046\times10^5$&$4.139\times10^5$\\
 $N_{re}$ 
 & $162$&$989$&$308$&$160$\\
 frac 
 & $0.00029$&$0.00028$&$0.00061$&$0.00038$
 \\

 $R_{\epsilon, \mu}$ \footnotetext[0]{$R_{X, \mu} = \frac{\overline{X_{\text{unconditional}}}}{\overline{X_{\text{conditional}}}}
$ and $R_{X, \sigma^2} = \frac{\text{variance}(X_{\text{unconditional}})}{\text{variance}(X_{\text{conditional}})}
$}&$0.577$&$0.581$&$0.474$& $0.362$\\

 $R_{\epsilon, \sigma^2}$ &$9.876$&$9.346$&$6.342$& $3.110$\\

 
$R_{T_{i}, \mu}$ &$0.083$&$0.083$&$0.093$& $0.110$\\

 $R_{T_{i}, \sigma^2}$ &$4.203$&$4.371$&$6.212$& $4.865$\\


 $ R_{E_{p}, \mu}$ &$0.932$&$0.083$&$0.933$& $0.915$\\

 $R_{E_{p}, \sigma^2}$ &$20.089$&$27.751$&$42.130$& $52.094$\\

\end{tabular}
\end{ruledtabular}
\end{table*}

\begin{table*}
\caption{\label{tab:table2}Forcing frequency 150 Hz}
\begin{ruledtabular}
\begin{tabular}{ccccccc}
 &\multicolumn{2}{c}{}&\multicolumn{2}{c}{}\\
   &0.5Vpp&1.0Vpp&2.0Vpp
&3.0Vpp&4.0Vpp&5.0Vpp\\ \hline
 $\eta_{\text{rms} (mm)}$&$0.083$&$0.338$ &$0.733$& $0.919$&$1.049$& $1.141$\\
 $V_{\text{rms,inj} (\frac{m}{s})}$ &$0.054$&$0.098$&$0.199$&$0.302$&$0.396$& $0.435$\\
 K&$3.089$&$3.098$&$2.994$&$3.025$&$3.015$& $3.021$\\
 S&$0.042$&$0.002$&$0.071$&$0.062$&$0.025$& $0.027$\\
 $N_{tot}$& $2.123\times10^6$&$9.270\times10^5$&$7.905\times10^5$&$9.166\times10^5$&$1.026\times10^6$& $1.027\times10^6$\\
 $N_{re}$& $15$&$206$&$315$&$323$&$366$& $337$\\
 frac& $7.064e-06$&$0.00022$&$0.00039$&$0.00035$&$0.00035$& $0.00032$\\

 $R_{\epsilon, \mu}$ &$0.785$&$0.797$&$0.642$& $0.666$&$0.702$& $0.711$\\

 $R_{\epsilon, \sigma^2}$ &$0.576$&$4.645$&$5.538$& $5.357$&$4.826$& $5.798$\\

 
$R_{T_{i}, \mu}$ &$0.064$&$0.042$&$0.055$& $0.052$&$0.049$& $0.049$\\

 $R_{T_{i}, \sigma^2}$ &$0.0221$&$0.225$&$1.271$& $1.258$&$0.939$& $0.993$\\


 $ R_{E_{p}, \mu}$ &$0.785$&$1.002$&$0.997$& $1.021$&$1.022$& $1.017$\\

 $R_{E_{p}, \sigma^2}$ &$6.478$&$50.940$&$65.450$& $48.088$&$68.801$& $78.989$\\

\end{tabular}
\end{ruledtabular}
\end{table*}
\begin{table*}
\caption{\label{tab:table3}Forcing frequency 250 Hz}
\begin{ruledtabular}
\begin{tabular}{ccccccccccc}
 &\multicolumn{2}{c}{}&\multicolumn{2}{c}{}\\
   &2.0Vpp\footnotemark[1]
&3.0Vpp\footnotemark[1]&3.2Vpp\footnotemark[1]&3.4Vpp\footnotemark[1]&3.6Vpp&3.8Vpp&4.0Vpp\footnotemark[2]&4.5Vpp&7.0Vpp\\ \hline
 $\eta_{\text{rms} (mm)}$ &$0.117$& $0.137$&$0.163$& $0.177$& $0.600$& $0.654$& $0.638$& $0.639$& $0.727$\\
 $V_{\text{rms,inj} (\frac{m}{s})}$ &$0.098$&$0.148$&$0.155$& $0.165$& $0.178$& $0.179$& $0.200$& $0.232$& $0.288$\\
 K&$2.990$&$3.007$&$3.030$& $3.089$& $3.154$& $3.090$& $3.077$& $2.955$& $2.976$\\
 S&$0.078$&$0.062$&$0.120$& $0.142$& $0.137$& $0.119$& $0.123$& $0.062$& $0.032$\\
$N_{tot}$&$1.673\times10^6$&$1.572\times10^6$&$1.910\times10^6$& $1.772\times10^6$& $1.158\times10^6$& $9.817\times10^5$& $6.980\times10^5$& $1.066\times10^6$& $1.124\times10^6$\\
$N_{re}$&$57$&$58$&$341$& $307$& $349$& $340$& $222$& $255$& $301$\\
 frac&$3.407\times10^{-5}$&$3.689\times10^{-5}$&$0.00017$& $0.00017$& $0.00030$& $0.00034$& $0.00031$& $0.00023$& $0.00026$\\
 
 $R_{\epsilon, \mu}$  &$0.693$&$0.766$&$1.002$& $0.998$& $0.797$& $0.705$& $0.781$& $0.765$& $0.759$\\

 $R_{\epsilon, \sigma^2}$ &$3.460$&$1.927$&$1.362$& $1.072$& $4.673$& $4.345$& $4.494$& $4.889$& $0.3758$\\

 
$R_{T_{i}, \mu}$ &$0.120$&$0.089$&$0.068$& $0.060$& $0.041$& $0.045$& $0.042$& $0.042
$& $0.042$\\

 $R_{T_{i}, \sigma^2}$ &$0.378$&$0.089$&$0.025$& $0.020$& $0.205$& $0.487$& $0.20$& $0.434$& $0.375$\\


 $R_{E_{p}, \mu}$ &$0.998$&$1.040$&$1.091$& $1.078$& $1.062$& $1.039$& $1.053$& $1.042$& $1.043$\\

 $R_{E_{p}, \sigma^2}$ &$21.456$&$28.249$&$23.651$& $24.469$& $97.773
$& $102.040$& $96.376$& $104.820$& $62.534$\\

\end{tabular}
\end{ruledtabular}
\footnotetext[1]{5-hour measurement}
\footnotetext[2]{6-hour measurement}
\end{table*}

\end{document}